\documentclass[conference]{IEEEtran}
\IEEEoverridecommandlockouts
\usepackage{mathptmx} 

\newcommand{\ignore}[1]{}
\usepackage{fancyhdr}
\usepackage[normalem]{ulem}
\usepackage[hyphens]{url}
\usepackage{microtype}

\usepackage{graphicx}
\usepackage{subfig}
\usepackage{authblk}
\graphicspath{{./figures/}}
\usepackage{url}
\usepackage{colortbl}
\definecolor{mygray}{gray}{.88}
\usepackage{multirow}
\usepackage{amsmath}
\usepackage{marvosym}
\usepackage{comment}

\newcommand{\tabincell}[2]{\begin{tabular}{@{}#1@{}}#2\end{tabular}}
\newcommand{\comp}{\hspace{0pt}\nolinebreak-\hspace{0pt}}

\usepackage[bookmarks=false]{hyperref}

\def\BibTeX{{\rm B\kern-.05em{\sc i\kern-.025em b}\kern-.08em
    T\kern-.1667em\lower.7ex\hbox{E}\kern-.125emX}}
\begin{document}

\title{Data Motif-based Proxy Benchmarks for Big Data and AI Workloads\\
}

\author[1,2]{Wanling Gao}
\author[1,2*]{Jianfeng Zhan \thanks{* The corresponding author is Jianfeng Zhan.}}
\author[1]{Lei Wang}
\author[1]{Chunjie Luo}
\author[3]{Zhen Jia}
\author[1]{Daoyi Zheng}
\author[1]{Chen Zheng}
\author[1]{\\Xiwen He}
\author[4]{Hainan Ye}
\author[5]{Haibin Wang}
\author[1]{Rui Ren}
\affil[1]{State Key Laboratory of Computer Architecture, Institute of Computing Technology, Chinese Academy of Sciences}
\affil[2]{University of Chinese Academy of Sciences, China}
\affil[3]{Princeton University}
\affil[4]{Beijing Academy of Frontier Sciences and Technology}
\affil[5]{Huawei}
\affil[ ]{\{gaowanling, zhanjianfeng, wanglei\_2011, luochunjie\}@ict.ac.cn, zhenj@princeton.edu, \{zhengdaoyi, zhengchen,}
\affil[ ]{hexiwen\}@ict.ac.cn, yehainan@mail.bafst.com, benjamin.wanghaibin@huawei.com, renrui@ict.ac.cn}


\maketitle

\begin{abstract}

For the architecture community, reasonable simulation time is a strong requirement in addition to performance data accuracy. However, emerging big data and AI workloads are too huge at binary size level and prohibitively expensive to run on cycle-accurate simulators. 
The concept of data motif, which is identified as a class of units of computation performed on initial or intermediate data, is the first step towards building proxy benchmark to mimic the real-world big data and AI workloads. However, there is no practical way to construct a proxy benchmark based on the data motifs to help simulation-based research.


In this paper, we embark on a study to bridge the gap between data motif and a practical proxy benchmark. We propose a data motif-based proxy benchmark generating methodology by means of machine learning method, which combine data motifs with different weights to mimic the big data and AI workloads. 
Furthermore, we implement various data motifs using light-weight stacks and apply the methodology to five real-world workloads to construct a suite of proxy benchmarks, considering the data types, patterns, and distributions. The evaluation results show that our proxy benchmarks shorten the execution time by 100s times on real systems while maintaining the average system and micro\comp architecture performance data accuracy above 90\%, even changing the input data sets or cluster configurations. Moreover, the generated proxy benchmarks reflect consistent performance trends across different architectures.
To facilitate the community, we will release the proxy benchmarks on the project homepage http://prof.ict.ac.cn/BigDataBench.

\end{abstract}


\section{Introduction}

Two concernful but interactional factors -- simulation accuracy and time, which always have trade-offs, decide the quality and time cycle of simulation-based research. 
Big data and AI workloads usually own thousands of billions of instructions because of the heavy software stack~\cite{jia_bigDataBench_subset} 
and long running time even on real machines. It is prohibitively expensive to run big data and AI workloads on cycle-accurate simulators, which will slow down on the order of hundred times\cite{simulatorSlow_Micro_04}.  

Hence, researchers from both academia and industry are committed to reducing the simulation time while keeping high accuracy. 
Simulation methods like sampled simulation~\cite{conte1996reducing,wunderlich2003smarts,sherwood2002automatically,lu2011efficient} and statistical simulation~\cite{skadron2003challenges,eeckhout2004control,nussbaum2001modeling,oskin2000hls} are proposed to generate synthetic trace or synthetic benchmarks and mimic micro-architecture performance of long-running real-world workloads, e.g., SPEC CPU~\cite{speccpu}. 
However, for emerging big data and AI workloads, which reveal significantly differential behaviors with traditional ones like SPEC CPU~\cite{speccpu} and PARSEC~\cite{bienia11benchmarking}, previous methods are challenged by the following limitations.
First, existing big data or AI benchmarks are difficult to be directly transplanted to simulators like GEM5~\cite{binkert2011gem5} because of the heavy software stacks and long running time. The distributed environment further aggravates this issue.
Second, the multithreaded workloads may exhibit different system behaviors in each run because of the nondeterminism~\cite{luo2005simulating,lepak2003redeeming,alameldeen2003variability,patil2004pinpointing}. Thus, the simulation points identified in one run using traditional simulation methods may not exist in another run~\cite{luo2005simulating,patil2004pinpointing}.
Moreover, synthetic traces or benchmarks target one workload on a specific architecture with certain configurations, and thus cannot accommodate other architectures or configurations~\cite{joshi2007constructing}.
Third, traditional methods fail to consider the impact of input data. While for big data and AI workloads, the input data has a great impact on workload behaviors~\cite{xie2018cvr,gao2018motif}, because of the diversities of data types, patterns and distributions.
In summary, a new benchmark methodology is urgently needed for emerging big data and AI workloads, which can generate proxy benchmarks satisfying the simulation time and accuracy requirements.





Data motifs~\cite{gao2018motif}, which are defined as the most time-consuming units of computation performed on different initial or intermediate data, are identified as one of the most promising key methods to solve the dilemma between simulation accuracy and runtime for big data and AI workloads.
Each data motif captures the common requirements while being reasonably divorced from individual implementations~\cite{asanovic2006landscape}, so
each big data or AI workload can be considered as a bunch of data motifs performed on different data. 
Even though motifs have been proposed to represent computation patterns of real-world workloads over a decade,  
there is still no work actually pulls it off and builds motif-based proxy benchmarks~\cite{colella2004defining,asanovic2006landscape}. This leaves us a gap between motifs and proxy benchmarks. Moreover, previous concept ``motifs" or kernels -- a set of operations extracted from original application~\cite{lilja2005measuring,hennessy2011computer}, cares more about the computation or communication patterns while paying little consideration on data types and patterns. However, the data motifs not only cover the algorithm diversity but also cover the data types, patterns and distributions, and thus reflect both the system and architecture characteristics. Considering the data diversity of big data and AI workloads, we construct proxy benchmarks based on ``data motifs".

To bridge the gap, we propose a data motif-based proxy benchmark generating methodology, in which we assemble a bunch of data motifs to mimic the behaviors of real-world big data or AI workloads.
Our methodology applies machine learning methods, which give us the opportunity to learn the inner connection between data motifs and real-world workloads, to reason about the components of proxy benchmark. 
In this paper, we use the decision tree as our first try to guide the generation of proxy benchmark. 
In the future, we will apply other advanced algorithms, like neural network.
To construct a proxy benchmark, we adopt a DAG-like structure, using a node 
to represent original or intermediate data set being processed, and an edge to represent a data motif.
Based on the methodology, we implement the big data motifs and AI data motifs individually using light-weight stacks to minimize the code binary size, 
and then generate big data and AI proxy benchmarks. 
Our current methodology and proxy benchmarks focus on average behaviors and ignore transient behaviors of the original benchmarks.  

Our contributions are three-fold as follows:

\begin{itemize}
\item We propose a data motif-based proxy benchmark generating methodology to bridge the gap between motifs and proxy benchmark. To the best of our knowledge, for the first time, we apply machine learning algorithm, i.e., decision tree, to the benchmark generating methodology, which enables the automatic synthesis of data motifs.
\item We provide five proxy benchmarks to represent five real-world big data and AI workloads. Our evaluations show the proxy benchmarks share high behavior similarities (above 90\% on average) with the real-world workloads, while shortening the execution time by 100s times. 
\item We use the five proxy benchmarks to perform three case studies, which further prove that, our methodology not only accommodates with different input data and configurations, but also reflects consistent performance trends across different architectures.
\end{itemize}

The rest of the paper is organized as follows.
Section 2 presents our data motif-based proxy benchmark generating methodology. Section 3 performs evaluations on a five-node X86\_64 cluster. In Section 4, we report three case studies.
Section 5 illustrates related work. Finally, we draw a conclusion in Section 6.

\section{Proxy Benchmark Generating Methodology}\label{tuning}

\begin{figure*}[!t]
\centering
\includegraphics*[scale=0.75]{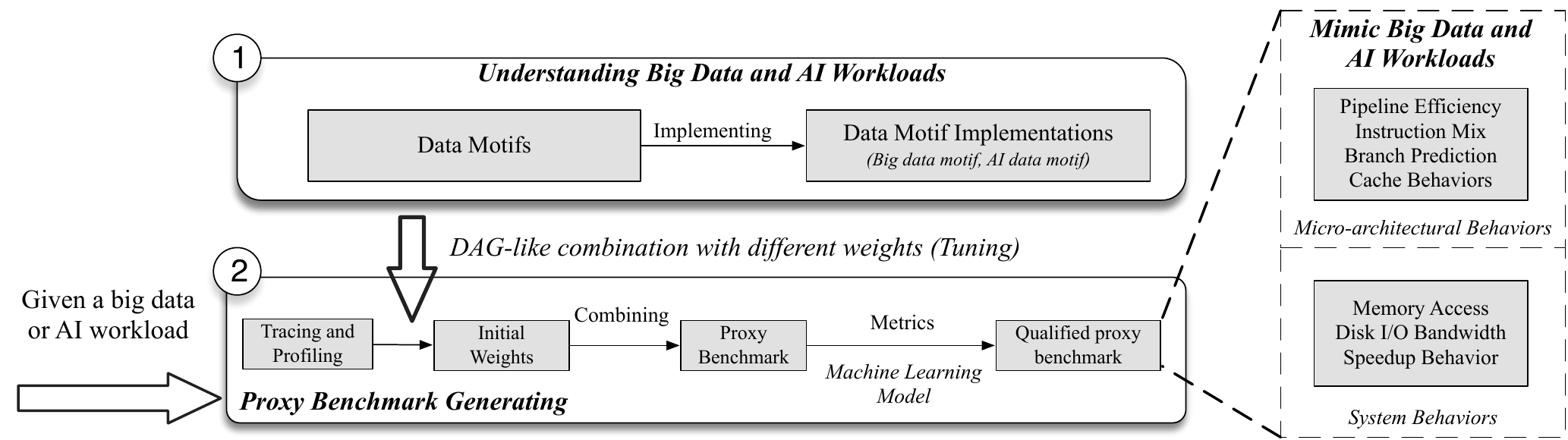}
\caption{Methodology Overview.} 
\label{motifidentify}
\end{figure*}

The data motif-based proxy benchmark generating methodology is illustrated in Fig.~\ref{motifidentify}. 
The whole methodology consists of the following steps.
For a given workload, we first get its system and architecture level profiles to identify the hotspot functions.
Then we correlate hotspots to the code fragments of the workload through bottom-up analysis. 
After that, we select the corresponding data motif implementations to represent the real-world workload.
The final proxy benchmarks are generated using a DAG-like combination of data motifs.
To generate qualified proxy benchmark that satisfies the requirements of performance data accuracy, such as cache behaviors or I/O behaviors, we provide an auto-tuning tool to tune the parameters of both data motif and the proxy benchmark.



In the following part of this section, we detail the proxy benchmark generating methodology.

\subsection{Eight Data Motifs} \label{component}

Previous work has identified eight data motifs~\cite{gao2018motif} as the most time-consuming units of computation performed on different initial or intermediate data, among a majority of big data and AI workloads. These eight data motifs are Matrix, Sampling, Transform, Graph, Logic, Set, Sort and Statistics. They are promising fundamental tools for benchmarking, designing, measuring, and optimizing big data and AI systems. 

Among them, matrix computation involves vector-vector, vector-matrix, and matrix-matrix computations. Sampling is a method to select a subset of original data according to a  certain statistical population.
Transform computation indicates the conversion from the original domain to another domain, such as fast fourier transform (FFT). Graph computation uses nodes representing entities and edges representing dependencies. Much previous work~\cite{beamer2015gap,beamer2015locality,dai2018graphh} focuses on graph workload evaluation, indicating its importance. 
Logic computation performs bit manipulation computations.
Set computation means the operations on one or more collections of distinct data, and also includes the primitive operators in relational algebra~\cite{codd1970relational}.
Sort and statistics are fundamental units of computation in big data and AI.

\begin{figure}[!t]
\centering
\includegraphics*[scale=0.61]{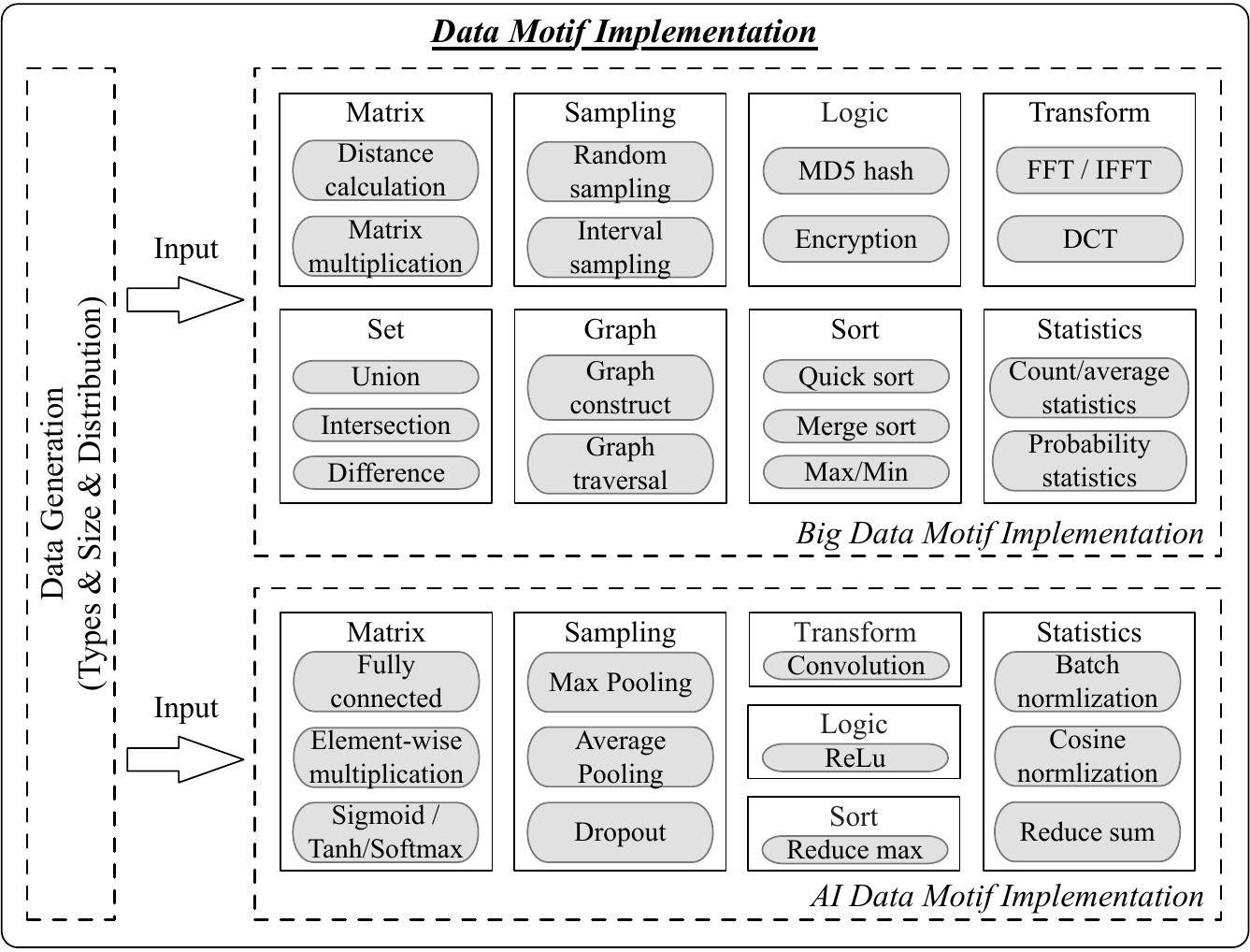}
\caption{The Overview of the Data Motif Implementations.} 
\label{library}
\end{figure}

Based on the eight data motifs, we provide light-weight implementations using POSIX threads model~\cite{butenhof1997programming}, as illustrated in Fig.~\ref{library}. 
Since the workload behaviors are sensitive to input data, we guarantee the sensitiveness of data motifs from the perspectives of data input and implementation. We provide various data input with different types, patterns and distributions, i.e., covering text, graph and matrix data, through data generation tools. Our implementation considers the execution model of software stacks and the programming styles of workloads using specific software stacks, which have great influences on workload behaviors~\cite{wang2014bigdatabench,jia_bigDataBench_subset}.
Fig.~\ref{library} lists all data motif implementations for both big data and AI. We use the POSIX threads model and consider the processes of original big data and AI software stacks. For example, we design the big data motif implementations from the perspectives of input data partition, chunk data allocation per thread, intermediate data written to disk, and data combination. 
In addition, big data systems like Hadoop adopt automatic memory management scheme, and thus incur many JVM garbage collection (GC) steps. So for big data motif implementations, we implement a unified memory management module, whose mechanism is similar with GC.
For AI data motif implementation, we consider the height size, width size and the number of channels of the input data or the convolution filter, the data storage format like the ``NHWC" or ``NCHW" in TensorFlow, the batch size, the stride of the sliding window, and the padding algorithm.

Different with kernels, our data motif implementations take real data set as input and have not only computation patterns and memory access patterns, but also disk I/O patterns.

\subsection{Proxy Benchmarks Construction}\label{tuning}

\begin{figure*}[!t]
\centering
\includegraphics*[scale=0.77]{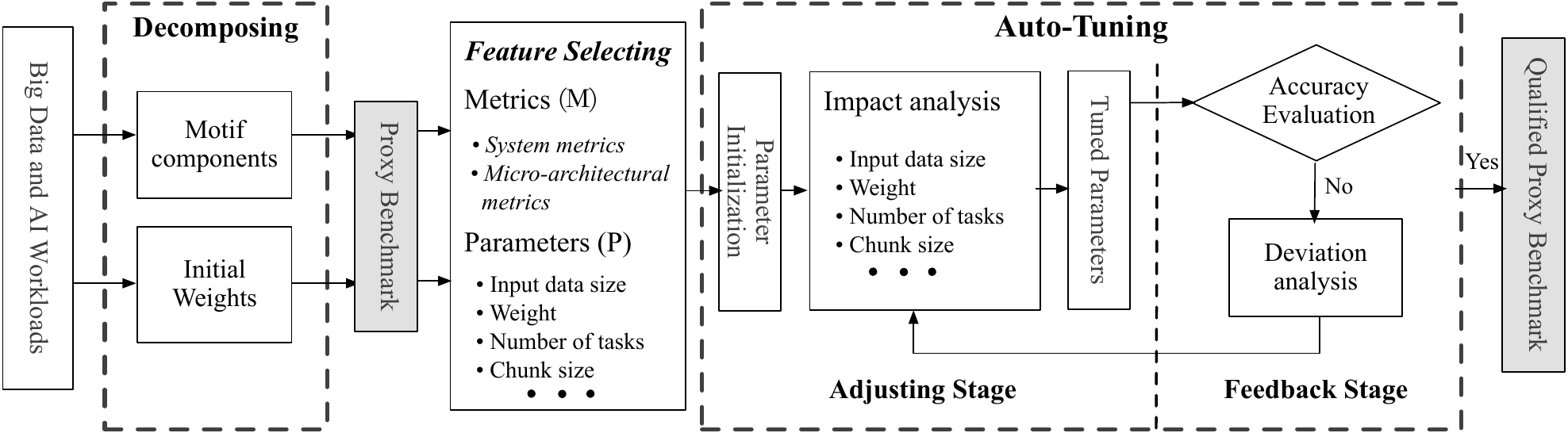}
\caption{Proxy Benchmarks Construction.} 
\label{tuning}
\end{figure*}

Fig.~\ref{tuning} presents the process of proxy benchmark construction, including \emph{decomposing} process, \emph{feature selecting} process, and \emph{tuning} process. We first break down the big data and AI benchmark into a group of data motifs and then tune them to approximate the original benchmark. We measure the proxy benchmark's accuracy by comparing the performance data of the proxy benchmark with those of the original workloads at both system and micro\comp architecture level.
To tune the accuracy---making it more similar to the original workload,  we further provide an auto-tuning tool using a machine learning model, decision tree.

\subsubsection{Benchmark Decomposing}
Given a big data or AI workload, we obtain its hotspot functions and execution time through a multi\comp dimensional tracing and profiling method, including runtime tracing (e.g. JVM tracing and logging), system profiling (e.g. CPU time breakdown), and hardware profiling (e.g. CPU cycle breakdown).
Based on the hotspot analysis, we correlate the hotspot functions to the code fragments of the workload and choose the corresponding data motif implementations by analyzing the computation logic of the code fragments.
Our proxy benchmark is a DAG-like combination of the selected data motifs with initial weights setting by their execution ratios. 




\subsubsection{Feature Selecting}
The main purpose of constructing proxy benchmarks is to mimic the system and micro-architectural behaviors of real-world workloads using data motif combinations.
The feature selecting stage is used to choose the concerned metrics and initialize the parameters of data motifs.

\textbf{System and Micro-architectural Metrics}
 According to the different concerns about the workloads, we can choose different metrics to tune a qualified proxy benchmark. For example, if our proxy benchmarks focus on cache behaviors of the workload, we can choose the  metrics that reflect cache behaviors like cache hit ratio to tune a qualified proxy benchmark.
Here we use \emph{$\overrightarrow{M}$} to denote the performance data of selected metrics. where:
$$ \overrightarrow{M} = (\;runtime, \;IPC, \;MIPS, \;L1D\; hitR, \;L2\; hitR, \;... ... ) $$
For system-level metrics, we choose running time, memory bandwidth, and disk I/O behavior.
For micro-architectural metrics, we choose instruction mix, cache behavior, branch prediction, and processor performance (i.e. IPC, MIPS).

\textbf{Parameters of Data Motifs}
We examine the configurable parameters of all data motifs and list them in Table~\ref{factor}.
Besides that, we further introduce \emph{weight} to indicate the contribution of each data motif.
For the rationality, we set the initial value of \emph{weight} proportional to their corresponding execution ratios. For example, in Hadoop TeraSort, the \emph{weight} is 70\% of sort computation, 10\% of sampling computation, and 20\% of graph computation, respectively. During the modeling process, the weight of each data motif can be adjusted within a reasonable range (e.g. plus or minus 10\%).

For better elaboration, we use $\overrightarrow{P}$ to indicate the parameter vector, which consists of all the configurable parameters.
When we discuss a specific data motif, only related elements in $\overrightarrow{P}$ will be concerned and others will be set to zero.
\begin{multline}
 \overrightarrow{P} = (\;dataSize, \;chunkSize, \;numTasks, \;weight \\
 \;batchSize, \;totalSize, \;heightSize, \;widthSize, \\
 \;numChannels)
\end{multline}

Each time we run the data motifs with a specific $\overrightarrow{P}$, there would be a resulted $\overrightarrow{M}$. 
To mimic the micro-architectural behavior of a real-world workload, we need to find the optimal $\overrightarrow{P}$ whose corresponding $\overrightarrow{M}$ is close enough to the metrics of this original workload.

 We initialize the $\overrightarrow{P}$ according to the configuration of the original workload.
We scale down the input data set and chunk size of the original workloads to initialize \emph{dataSize} and \emph{chunkSize}.
The \emph{numTasks} is initialized as the parallelism degree of the original workload.

\begin{table}[htb]
\caption{Tunable Parameters for Each Data Motif.}\label{factor}
\renewcommand\arraystretch{1.2}
\center
\footnotesize
\begin{tabular}{|p{0.6in}|p{2.2in}|}
  \hline
  \textbf{Parameter} & \textbf{Description} \\
  \hline 
  \emph{dataSize} & The input data size for each big data motif \\
  \hline
  \emph{chunkSize} & The data block size processed by each thread for each big data motif \\
  \hline
  \emph{numTasks} & The process and thread numbers for each big data and AI data motif \\
  \hline
  \emph{batchSize} & The batch size of each iteration for each AI data motif \\
  \hline
  \emph{totalSize} & The total input data size need to be processed for each AI data motif\\
  \hline
  \emph{heightSize} & The height dimension for one input data or filter \\
  \hline
  \emph{widthSize} & The width dimension for one input data or filter \\
  \hline
  \emph{numChannels} & The channel number for one input data or filter \\
  \hline
  \emph{weight} & The contribution for each data motif \\
  \hline
\end{tabular}
\end{table}

%
%
%
\subsubsection{Adjusting Stage}
We introduce a decision tree based mechanism to assist the auto-tuning process.
The tool learns the impact that each parameter in $\overrightarrow{P}$ will have on $\overrightarrow{M}$ and builds a decision tree through \emph{Impact analysis}. 
The learning process changes one parameter each time and execute multiple times to characterize the parameter's impact on each metric.
Based on the impact analysis, the tool builds a decision tree to determine which parameter to tune if one metric has a  large deviation.
After that, the \emph{Feedback Stage} is activated to evaluate the proxy benchmark with tuned parameters. If it does not satisfy the requirements, the tool will adjust the parameters to improve the accuracy using the decision tree in the adjusting stage.


\subsubsection{Feedback Stage}
In the feedback stage, the tool evaluates the accuracy of the current proxy benchmark with specific parameters.
If the deviations of all metrics are within the setting range (e.g. 15\%), the auto-tuning process is finished.
Otherwise, the metrics with large deviations will be fed back to the adjusting stage.
The adjusting and feedback processes will iterate until reaching the specified accuracy, and the finalized proxy benchmark with the final parameter settings is our qualified proxy benchmark. Sampling performance counters may have slight variations when performing multiple runs, so the parameter values of qualified proxy benchmark may also have slight variations. However, the deviations of all metrics are within the setting range according to the end conditions.



\subsection{Discussion}

\begin{table*}[htbp]
\centering
\caption{Comparison of Different Simulation Methodologies for Big Data and AI Workloads.}\label{simulation-compare}
\center
\begin{tabular}{|p{1.2in}|p{0.9in}|p{0.6in}|p{0.7in}|p{0.65in}|p{0.6in}|p{0.5in}|}
\hline
\multirow{2}{*}{\tabincell{c}{Methodology}} & \multirow{2}{*}{\tabincell{c}{Typical\\Benchmark or Tool}} &\multirow{2}{*}{\tabincell{c}{ Data Set}}
& \multirow{2}{*}{\tabincell{c}{Portable Cost}}
&\multirow{2}{*}{\tabincell{c}{Multi-core\\Scalability}}&\multirow{2}{*}{\tabincell{c}{Cross \\ Architecture}}&\multirow{2}{*}{\tabincell{c}{Accuracy}}\\
& & & & & &\\
\hline
\tabincell{c}{Kernel Benchmark}& NPB~\cite{bailey1991parallel} & Fixed & Recompile & Yes & Yes & Low  \\
\hline
\tabincell{c}{Synthetic Trace Method} & SimPoint~\cite{hamerly2005simpoint} & Fixed & Regenerate & No & No & High  \\
\hline
\tabincell{c}{Synthetic Benchmark}& PerfProx~\cite{panda2017proxypact} & Fixed & Regenerate & No & No & High \\
\hline
\tabincell{l}{Data Motif-Based\\Proxy Benchmark}& \tabincell{l}{Data Motif\\Benchmark} & On-demand & Recompile & Yes & Yes & High\\
\hline
\end{tabular}
\end{table*}
Table.~\ref{simulation-compare} compares four simulation methodologies from the perspectives of data set, portable cost,  multi-core scalability, cross architecture, and accuracy.

Kernel benchmarks, which consist of a set of kernels extracted from original application~\cite{lilja2005measuring,hennessy2011computer}, are widely used in high performance computing. However, they are insufficient to completely reflect workload behaviors of big data and AI workloads~\cite{bailey1991parallel,lilja2005measuring}. 
  

However, the real trace or statistical profile are obtained with the aid of a functional simulator or a binary instrumentation tool (e.g. Pin~\cite{luk2005pin,reddi2004pin}), which is time-consuming and costly. Complex big data or AI software stacks and their distributed deployments further aggravate this challenge. 
So generating synthetic trace is infeasible and time-consuming, especially for different architectures or workload configurations~\cite{eeckhout2000performance}. For example, previous work uses Pin and SimPoint to generate synthetic traces. However, Pin lacks support for diverse architectures (e.g. ARM architecture) and Java environment~\cite{panda2017proxypact}. So it is difficult to use Pin in big data systems like Hadoop. Another method is to use functional simulator (e.g. GEM5~\cite{binkert2011gem5}) and SimPoint to obtain traces. GEM5 has limited supports for distributed deployment and also takes a long time.

Synthetic benchmark is to generate assembly code or C code based on workload profiling~\cite{van2010benchmark}, and can work on real 
hardware as well as execution-driven simulators. However, existing synthetic benchmarks can only be used to mimic micro-architectural metrics, and do not support multi-thread model, and their codes do not contain computation logic.  So they can not be used to characterize system-level behaviors such as multi-core scalability or carry out cross architecture comparisons. Moreover, synthetic benchmarks need to be re-generated on different architectures or workload configurations.

In our data motif-based benchmarking method, we use multi-thread programs and preserve computation logic to mimic the behaviors of big data and AI workloads. Our proxy benchmarks can suit for different data input and support cross-architecture comparison with recompilation. 
As for simulation accuracy, they can reflect not only micro-architectural behaviors but also system-level behaviors of real-world big data and AI workloads.

\section{Evaluation}\label{evaluation}


Simulation time and performance data accuracy are mandatory requirements for architecture community. 
In this section, we evaluate the effectiveness of our proxy benchmark generating methodology through constructing five proxy benchmarks for big data and AI workloads. Then we measure these proxy benchmarks from the perspectives of runtime speedup and accuracy.

\subsection{Real-world Workloads and Proxy Benchmarks}

In order to cover different application domains and workload patterns, 
we choose five representative big data and AI workloads from BigDataBench 4.0~\cite{wang2014bigdatabench} -- Hadoop TeraSort, Hadoop K-means, Hadoop PageRank, TensorFlow AlexNet and TensorFlow Inception-V3. We choose them for the following reasons.

\textbf{Representative Application Domains} They are all widely used in many important application domains. For example, TeraSort is a widely-used workload in many application domains; PageRank is a famous workload for search engine; K-means is a simple but useful workload in internet services and machine learning community; 
AlexNet~\cite{krizhevsky2012imagenet} and Inception-V3~\cite{szegedy2016rethinking} are two typical convolutional neural networks (CNN) in artificial intelligence with different neural network structures.

\textbf{Various Workload Patterns} They have different workload patterns. Hadoop TeraSort is an I/O-intensive workload; Hadoop K-means is a CPU-intensive and memory-intensive workload; Hadoop PageRank is a both CPU-intensive and I/O-intensive workload; 
TensorFlow AlexNet is both CPU-intensive and memory-intensive, while Inception-V3 is a CPU-intensive workload.

\textbf{Diverse Data Inputs} They take different data as  inputs. Hadoop TeraSort uses text data generated by gensort~\cite{gensort}; Hadoop K-means uses vector data while Hadoop PageRank uses graph data; 
TensorFlow AlexNet uses image data from CIFAR-10~\cite{krizhevsky2009learning} or matrix data; TensorFlow Inception-V3 uses ILSVRC2012~\cite{russakovsky2012imagenet} image data. These benchmarks are of great significance for measuring big data and AI systems and architectures~\cite{jia_bigDataBench_subset}.

Also, we generate the corresponding five proxy benchmarks using our methodology. 
In the rest of this paper, we use Proxy TeraSort, Proxy K-means, Proxy PageRank, Proxy AlexNet, and Proxy Inception-V3 to represent the proxy benchmark for Hadoop TeraSort, Hadoop K-means, Hadoop PageRank, TensorFlow AlexNet, and TensorFlow Inception-V3, respectively.
Note that the input data to each proxy benchmark has the same data type and distribution with respect to those of the original big data and AI workloads, so as to preserve the impact of data on workload behaviors.

Table.~\ref{data_workloads} lists the benchmark details from the perspectives of workload pattern, data set, involved data motifs, and data motif implementations of the proxy benchmarks.

\begin{table*}[htbp]
\caption{Five Real Benchmarks and Their Corresponding Proxy Benchmarks.}\label{data_workloads}
\renewcommand\arraystretch{1.2}
\center
\footnotesize
\begin{tabular}{|p{0.81in}|p{1in}|p{0.6in}|p{1.2in}|p{2.52in}|}
  \hline
  \tabincell{l}{Big Data \& AI\\Benchmark} & \tabincell{l}{Workload Pattern} & \tabincell{l}{Data Set} & \tabincell{l}{Involved Data Motifs} & Data Motif Implementations of Proxy Benchmark  \\
  \hline
  \tabincell{l}{Hadoop\\TeraSort} &\tabincell{l}{I/O Intensive} & \tabincell{l}{Text} & \tabincell{l}{Sort\\Sampling\\Graph} & \tabincell{l}{Quick sort; Merge sort\\Random sampling; Interval sampling\\Graph construction; Graph traversal} \\
  \cline{1-5}
  \tabincell{l}{Hadoop\\K-means} &\tabincell{l}{CPU Intensive\\Memory Intensive} & \tabincell{l}{Vectors} & \tabincell{l}{Matrix\\Sort\\Statistics} & \tabincell{l}{Vector euclidean distance; Cosine distance\\Quick sort; Merge sort\\Cluster count; Average computation}  \\
  \cline{1-5}
  \tabincell{l}{Hadoop\\PageRank}&\tabincell{l}{CPU Intensive\\I/O Intensive} &\tabincell{l}{Graph} & \tabincell{l}{Matrix\\Sort\\Statistics } & \tabincell{l}{Matrix construction; Matrix multiplication\\Quick sort; Min/max calculation\\Out degree and in degree count of nodes} \\
  \hline
  \tabincell{l}{TensorFlow\\AlexNet} &\tabincell{l}{CPU Intensive\\Memory Intensive}&\tabincell{l}{Image/Matrix} & \tabincell{l}{Matrix \\Sampling\\Transform\\Statistics} & \tabincell{l}{Fully connected\\Max Pooling\\Convolution\\Batch normalization} \\
  \hline
  \tabincell{l}{TensorFlow\\Inception-V3} &\tabincell{l}{CPU Intensive}&\tabincell{l}{Image/Matrix} & \tabincell{l}{Matrix \\Sampling\\Logic\\Transform\\Statistics} & \tabincell{l}{Fully connected; Softmax\\Max pooling; Average pooling; Dropout \\ReLu\\Convolution\\Batch normalization} \\
  \hline
  
\end{tabular}
\end{table*}

\subsection{Experiment Setups}


We deploy a five-node cluster, with one master node and four slave nodes. They are connected using 1Gb ethernet network. Each node is equipped with two Intel Xeon E5645 (Westmere) processors, and each processor has six physical out-of-order cores. The memory of each node is 32GB, DDR3. Each node runs Linux CentOS 6.4 with the Linux kernel version 3.11.10. The JDK and Hadoop versions are 1.7.0 and 2.7.1, respectively. The GCC version is 4.8.0.
The proxy benchmarks are compiled using ``-O2" or ``-O3" option for optimization.
The hardware and software details are listed on Table \ref{hwconfigeration}.


To evaluate the performance data accuracy, we run the proxy benchmarks against the original big data and AI workloads. The Hadoop benchmarks are run on the above five-node cluster using the optimized Hadoop configurations, through tuning the data block size of the Hadoop distributed file system, memory allocation for each map/reduce job, reduce job numbers, and memory size according to the cluster scales.
For Hadoop TeraSort, we choose 100 GB text data produced by gensort~\cite{gensort}. For Hadoop K-means and PageRank, we choose 100 GB sparse vector data with 90\% sparsity~\footnote{The sparsity of the vector indicates the proportion of zero-valued elements.} and $2^{26}$-vertex graph both generated by BDGS~\cite{ming2014bdgs}, respectively. 

For AI benchmarks, we run the TensorFlow workloads on the above five-node cluster, with one node as parameter server and the other four as workers. The TensorFlow AlexNet workload uses the CIFAR-10~\cite{krizhevsky2009learning} dataset as the input and runs 10,000 steps in total with each worker running 2500 steps. The TensorFlow Inception-V3 workload uses the ILSVRC2012~\cite{russakovsky2012imagenet} image dataset as input and runs 1000 steps in total with each worker running 250 steps. The batch size is 128 and 32 for AlexNet and Inception-V3, respectively, considering the image size. 
For comparison, we run all five proxy benchmarks on one of the slave nodes, respectively.


\begin{table}
\caption{Node Configuration Details of Xeon E5645}\label{hwconfigeration}
\renewcommand\arraystretch{1.2}
\center
\footnotesize
\begin{tabular}{|p{0.7in}|p{0.66in}|p{0.66in}|p{0.66in}|}
\hline \rowcolor{mygray} \multicolumn{4}{|l|}{Hardware Configurations}\\
\hline \multicolumn{2}{|c|}{CPU Type} & \multicolumn{2}{c|}{Intel CPU Core} \\
\hline \multicolumn{2}{|c|}{Intel \textregistered Xeon E5645}  &\multicolumn{2}{c|}{6 cores@2.40G} \\
\hline L1 DCache &L1 ICache &L2 Cache &L3 Cache \\
\hline 6 $\times$ 32 KB& 6 $\times$ 32 KB&6 $\times$ 256 KB& 12MB \\
\hline \multicolumn{2}{|c|}{Hyper-Threading} & \multicolumn{2}{c|}{Disabled}\\
\hline
\end{tabular}
\end{table}


\subsection{Metrics Selection and Collection}\label{metricCol}

To evaluate the accuracy, we choose micro-architectural and system metrics covering instruction mix, cache behavior, branch prediction, processor performance, memory bandwidth and disk I/O behavior.
Table \ref{simulation_metric} presents the metrics we choose.


\textbf{Processor Performance}. We choose two metrics to measure the processor overall performance. Instructions per cycle (IPC) indicates the average number of instructions executed per clock cycle. Million instructions per second (MIPS) indicates the instruction execution speed.

\textbf{Instruction Mix}. We consider the instruction mix breakdown including the percentage of integer instructions, floating-point instructions, load instructions, store instructions and branch instructions.

\textbf{Branch Prediction}. Branch predication is an important strategy used in modern processors. We track the miss prediction ratio of branch instructions (br\_miss for short).

\textbf{Cache Behavior}. We evaluate cache efficiency using cache hit ratios, including L1 instruction cache, L1 data cache, L2 cache and L3 cache.

\textbf{Memory Bandwidth}. We measure the data load rate from memory and the data store rate into memory, with the unit of bytes per second. We choose metrics of memory read bandwidth (read\_bw for short), memory write bandwidth (write\_bw for short) and total memory bandwidth including both read and write (mem\_bw for short).

\textbf{Disk I/O Behavior}. We employ disk I/O bandwidth to reflect the I/O behaviors of the workloads. The disk I/O bandwidth is calculated by Equation \ref{io-bandwidth}, where $Sector_{Read+Write}$ means the total number of sector reads and sector writes; $Size_{Sector}$ means the sector size (512 bytes for our nodes).

\begin{equation} \label{io-bandwidth}
\begin{aligned}
BW_{Disk I/O}=\frac{(Sector_{Read+Write})*Size_{Sector}}{RunTime}
\end{aligned}
\end{equation}



We collect micro-architectural metrics from hardware performance monitoring counters (PMCs),
and look up the hardware events' value on Intel Developer's Manual~\cite{intel2010intel}. Perf~\cite{perftool} is used to collect these hardware events. To guarantee the accuracy and validity, we run each workload three times, and collect performance data of workloads on all slave nodes during the whole runtime. We report and analyze their average value.

\begin{table}[htb]
\caption{System and Micro-architectural Metrics.}
\renewcommand\arraystretch{1.1}
\center
\footnotesize
\begin{tabular}{|p{0.73in}|p{0.8in}|p{1.4in}|}
  \hline
  Category& Metric Name &  Description \\
  \hline \rowcolor{mygray} \multicolumn{3}{|l|}{Micro-architectural Metrics}\\
  \hline
  \multirow{2}{*}{\tabincell{l}{Processor\\Performance}} & IPC & \tabincell{l}{Instructions per cycle} \\
  \cline{2-3}
   & MIPS & \tabincell{l}{Million instructions per second} \\
  \hline
  \multirow{3}{*}{\tabincell{l}{Instruction\\Mix}} & \multirow{3}{*}{\tabincell{l}{Instruction\\ratios}} & \multirow{3}{*}{\tabincell{l}{Ratios of load, store, branch,\\ floating-point, and integer\\instructions}}  \\
  &&\\
  &&\\
  \hline
  \tabincell{l}{Branch Prediction}  & Branch Miss & Branch miss prediction ratio \\
  \hline
  \multirow{4}{*}{\tabincell{l}{Cache\\Behavior}}  & L1I Hit Ratio & L1 instruction cache hit ratio \\
  \cline{2-3}
   & L1D Hit Ratio & L1 data cache hit ratio \\
  \cline{2-3}
   & L2 Hit Ratio & L2 cache hit ratio \\
  \cline{2-3}
   & L3 Hit Ratio & L3 cache hit ratio \\
  \hline
  \rowcolor{mygray} \multicolumn{3}{|l|}{System Metrics}\\ \hline
  \multirow{4}{*}{\tabincell{l}{Memory\\Bandwidth}}  & \tabincell{l}{Read Bandwidth} & \tabincell{l}{Memory load bandwidth}\\
  \cline{2-3}
   & \tabincell{l}{Write Bandwidth} & \tabincell{l}{Memory store bandwidth}\\
  \cline{2-3}
   & \tabincell{l}{Total Bandwidth} & memory load and store bandwidth \\
  \hline
  \multirow{2}{*}{\tabincell{l}{Disk I/O\\Behavior}} &  \multirow{2}{*}{\tabincell{l}{Disk I/O\\Bandwidth}} & \multirow{2}{*}{Disk read and write bandwidth} \\
  &&\\
  \hline
\end{tabular}\label{simulation_metric}
\end{table}

\subsection{Runtime Speedup}

Table~\ref{time-e5645} presents the execution time of the real benchmarks and the proxy benchmarks on Xeon E5645. For big data benchmarks, Hadoop TeraSort with 100 GB text data runs 1500 seconds on the five-node cluster. Hadoop K-means with 100 GB vectors runs 5971 seconds for each iteration. Hadoop PageRank with $2^{26}$-vertex graph runs 1444 seconds for each iteration. 
For AI benchmarks, TensorFlow AlexNet with CIFAR-10 dataset runs 1556 seconds. TensorFlow Inception-V3 with ILSVRC2012 dataset runs 6782 seconds.
The five corresponding proxy benchmarks run about ten seconds on the physical machine. For TeraSort, K-means, PageRank, AlexNet, and Inception-V3, the speedup is 136X (1500/11.02), 743X (5971/8.03), 160X (1444/9.03), 155X (1556/10.02), and 376X (6782/18) respectively.

\begin{table}[htb]
\caption{Execution Time on Xeon E5645.}\label{time-e5645}
\renewcommand\arraystretch{1.2}
\center
\footnotesize
\begin{tabular}{|p{0.72in}|p{0.98in}|p{0.9in}|}
  \hline
  \multirow{2}{*}{Workloads} & \multicolumn{2}{c|}{Execution Time (Second)} \\
  \cline{2-3}
  & Real version & Proxy version \\
  \hline
  TeraSort & 1500 & 11.02 \\
  \hline
  K-means & 5971 & 8.03 \\
  \hline
  PageRank & 1444 & 9.03 \\
  \hline
  AlexNet & 1556 & 10.02 \\
  \hline
  Inception-V3 & 6782 & 18 \\
  \hline
\end{tabular}
\end{table}

\subsection{Accuracy}


We evaluate the accuracy of proxy benchmarks from system and micro-architecture perspectives, using all metrics listed in Table \ref{simulation_metric}. 

\textbf{System and Micro-architecture Data Accuracy.}
For each metric in Table~\ref{simulation_metric}, the accuracy of the proxy benchmark comparing to the real benchmark is computed by Equation \ref{similarity-equ1}. Among which, $Val_R$ represents the average value of the real benchmark on all slave nodes; $Val_P$ represents the average value of the proxy benchmark on a slave node. The absolute value ranges from 0 to 1. The number closer to 1 indicates higher accuracy.

\begin{equation} \label{similarity-equ1}
\begin{aligned}
Accuracy(Val_R,Val_P)=1-\left|\frac{Val_P-Val_R}{Val_R}\right|
\end{aligned}
\end{equation}

\begin{figure}[htb]
\centering
\includegraphics*[scale=0.6]{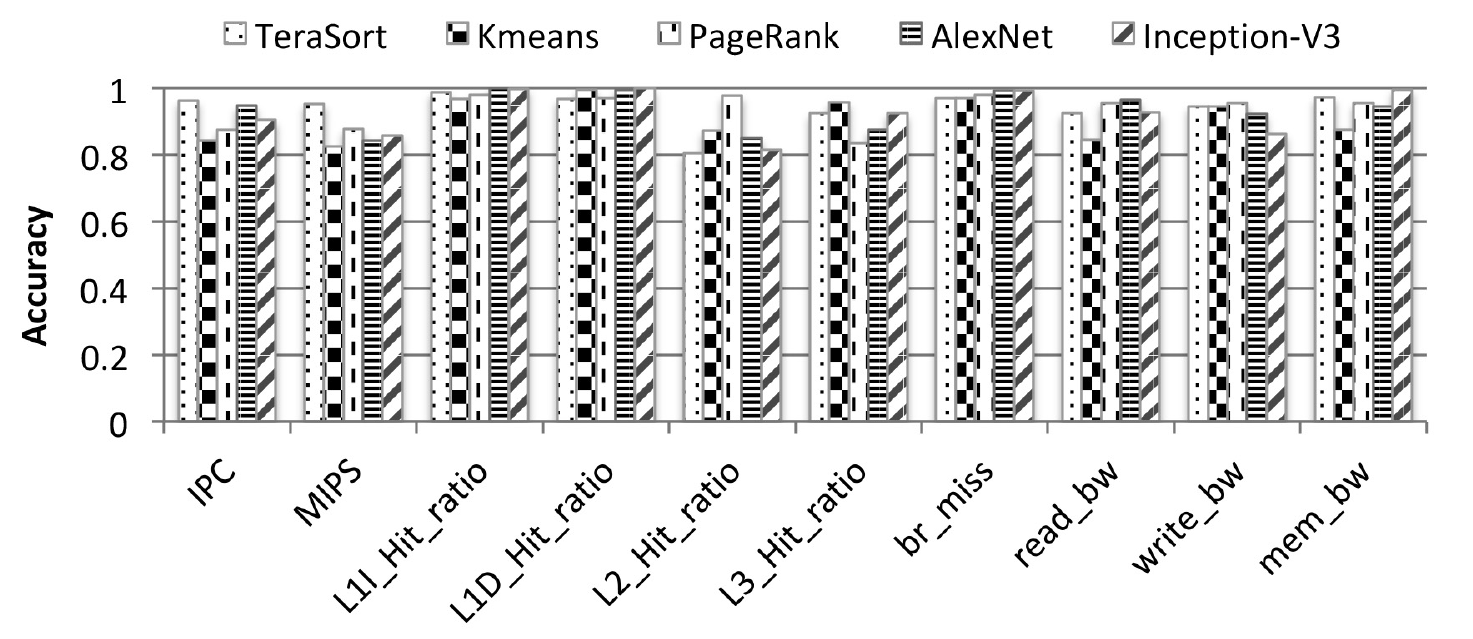}
\caption{System and Micro-architectural Data Accuracy on Xeon E5645.}
\label{e5645:metric}
\end{figure}

Fig.~\ref{e5645:metric} presents the system and micro-architectural data accuracy of the proxy benchmarks on Xeon E5645. We can find that the average accuracy of all metrics are greater than 90\%. For TeraSort, K-means, PageRank, AlexNet, and Inception-V3, the average accuracy is 94\%, 91\%, 93\%, 93.7\% and 92.6\%, respectively.

\begin{figure}[htb]
\centering
\includegraphics*[scale=0.54]{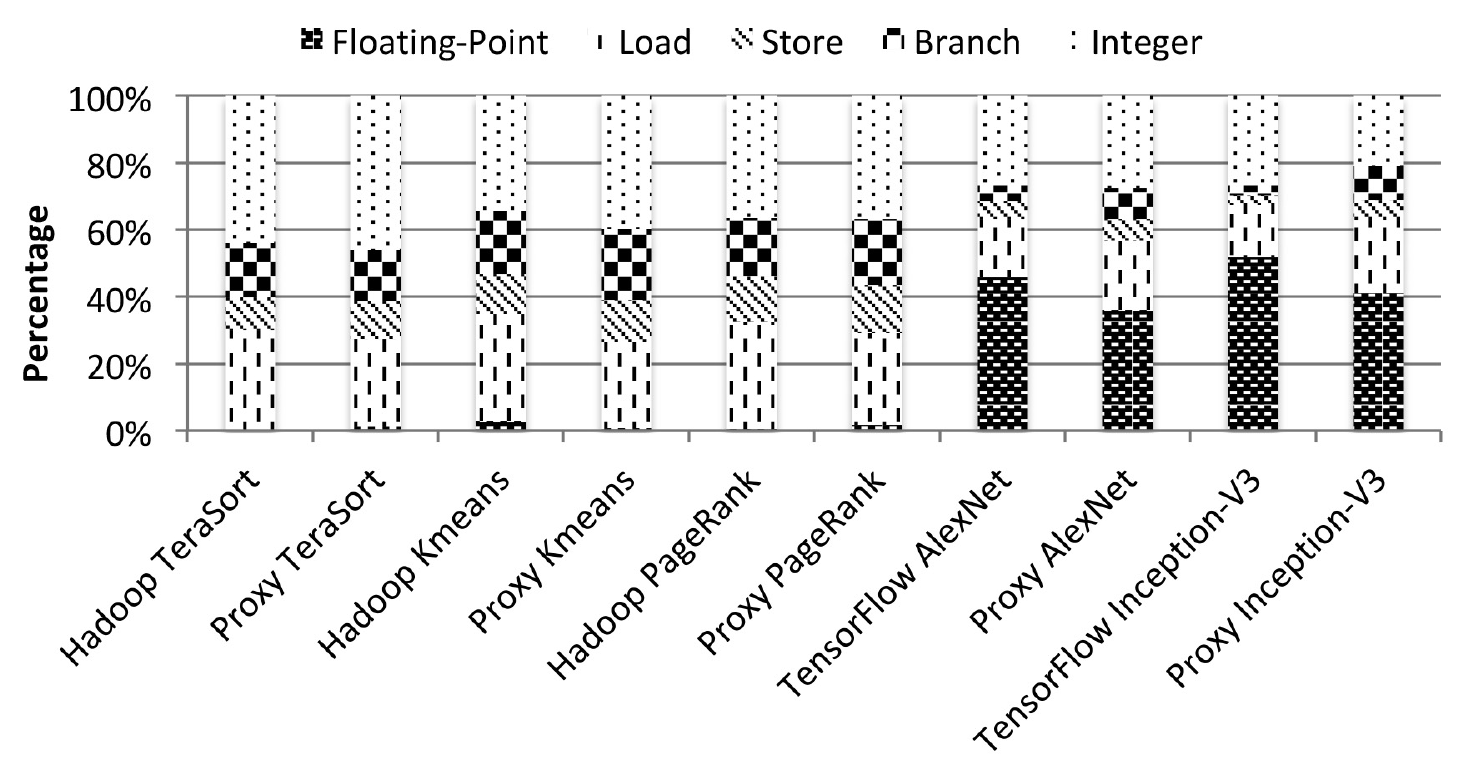}
\caption{Instruction Mix Breakdown on Xeon E5645.}
\label{e5645:inst}
\end{figure}

\textbf{Instruction Mix Breakdown.} 
Fig.~\ref{e5645:inst} shows the instruction mix breakdown of the proxy benchmarks and real benchmarks.
From Fig.~\ref{e5645:inst}, we can find that the five proxy benchmarks preserve the instruction mix characteristics of these five real benchmarks with Hadoop or TensorFlow stacks. For example, the integer instruction occupies 44\% for Hadoop TeraSort and 46\% for Proxy TeraSort, while the floating-point instruction occupies less than 1\% for both Hadoop and Proxy TeraSort. For instructions involving data movement, Hadoop TeraSort contains 39\% of load and store instructions, and Proxy TeraSort contains 37\%.   We find that TensorFlow workloads have much higher floating-point instructions than big data workloads, and our proxy benchmarks for TensorFlow workloads also reflect similar instruction mix breakdown behaviors. For example, TensorFlow AlexNet and Proxy AlexNet both contain a large percentage of floating-point instructions, about 40\%.

\begin{figure}[htb]
\centering
\includegraphics*[scale=0.6]{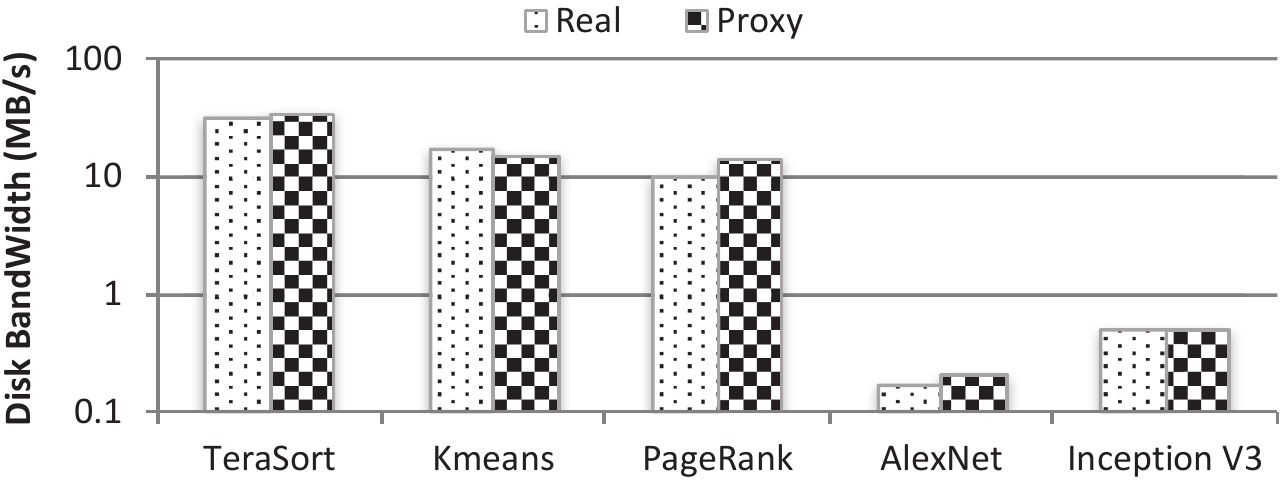}
\caption{Disk I/O Bandwidth on Xeon E5645.}
\label{e5645:io}
\end{figure}

\textbf{Disk I/O Behaviors.}
Big data applications have significant disk I/O pressures. 
We use Equation \ref{io-bandwidth} as illustrated in Subsection~\ref{metricCol} to compute the disk I/O bandwidth.
Fig.~\ref{e5645:io} presents the I/O bandwidth of proxy benchmarks and real benchmarks on Xeon E5645. We find that they have similar average disk I/O pressure. The disk I/O bandwidth of Proxy TeraSort and Hadoop TeraSort is 32.04 MB and 33.99 MB per second, respectively. 
However, for AI workloads, we find that they have extremely low disk I/O bandwidth, about 0.2 MB/s for AlexNet and 0.5 MB/s for Inception-V3. This is because that the deep learning workloads are CPU intensive and process a batch of input data every time, so the disk I/O pressure is not that much.

\section{Case Studies}


In this section, we would like to examine whether the generated proxy benchmark can retain the key properties of real-world big data and AI workloads.
To be specific, we will examine the proxy benchmarks from the following perspectives:
1). can proxy benchmark reflect the impact of input data.
2). is proxy benchmark adaptable to system level configuration. 
3). can proxy benchmark reflect the relative performance among different architectures.





\subsection{Data Input}

The data type and  access pattern have a great impact on the performance and behaviors for big data workloads. Among them, input data sparsity is one of the most important factors, which widely affects the design and decision in many fields, like recommendation systems and graph computing. 

\begin{figure}[htb]
\centering
\includegraphics*[scale=0.7]{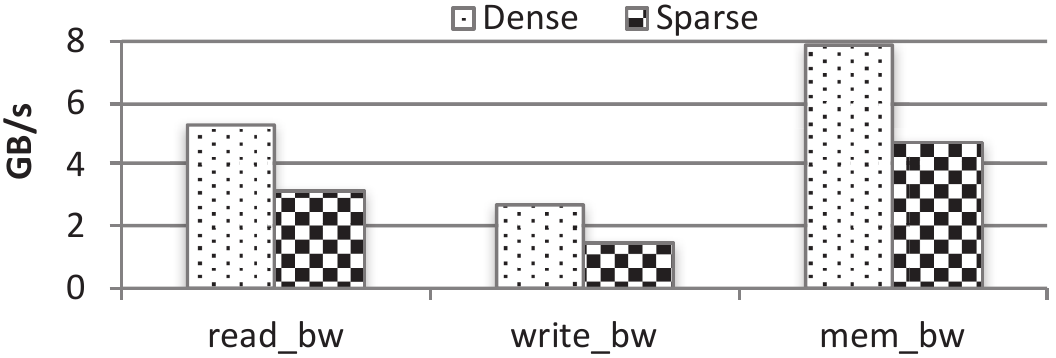}
\caption{Data Impact on Memory Bandwidth Using Sparse and Dense Data for Hadoop K-means on Xeon E5645.}
\label{e5645:sparse:diff}
\end{figure}

Among the five real workloads, K-means' behavior is highly affected by the sparsity of input data.
We use two input data sets with different sparsity to drive Hadoop K-means: sparse vector (the original configuration, 90\% elements are zero) and dense vectors (all elements are non-zero, and 0\% elements are zero). 
Fig.~\ref{e5645:sparse:diff} presents the memory bandwidth we measured by running Hadoop K-means with two different input datasets. We find that the memory bandwidth measured with sparse vectors is nearly half of that with dense vectors, which indicates that the sparsity of input data heavily affect the workload's behavior.

Then here, we want to check whether the behavior accuracy of the generated proxy benchmark is affected by the sparsity of input data. 
If the answer is no, which is ideal, we only need to generate one proxy benchmark for Hadoop K-means, and use it with different sparsity input data set. If yes, which is the case of synthetic trace method, we need to generate different proxy benchmarks for diverse sparsity input data sets.   

\begin{figure}[htb]
\centering
\includegraphics*[scale=0.68]{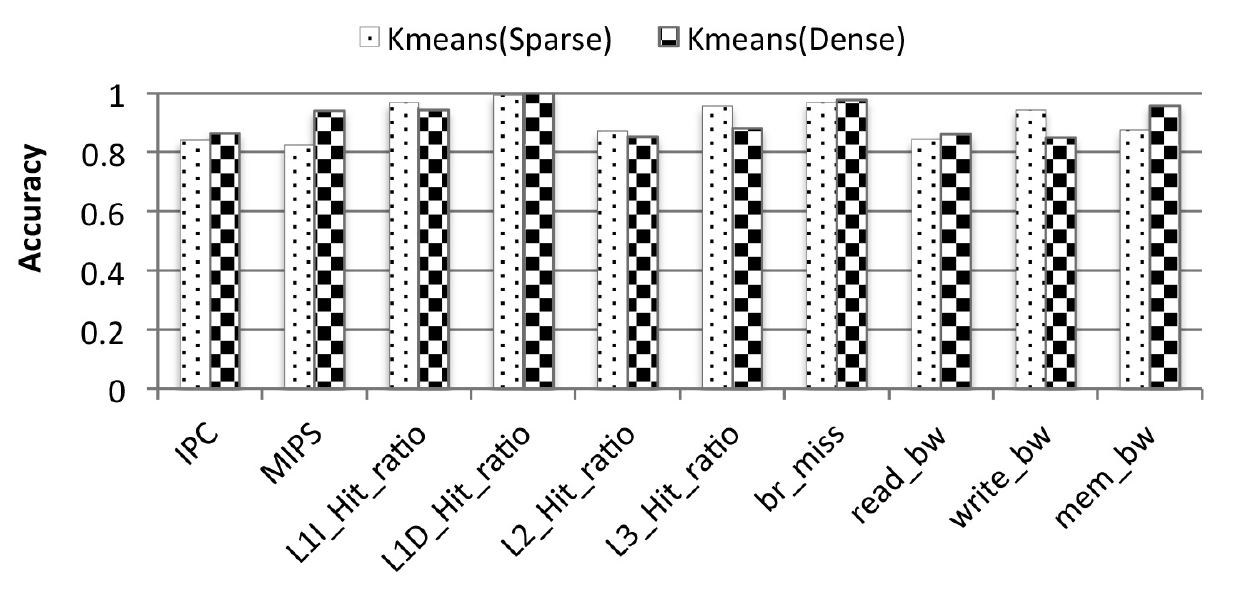}
\caption{System and Micro-architectural Data Accuracy Using Different Input Data on Xeon E5645.}
\label{e5645:dense:metric}
\end{figure}

Here we only generate one proxy benchmark for Hadoop K-means but drive it by two different sparsity data sets and compare the results with real Hadoop K-means.
Fig.~\ref{e5645:dense:metric} shows the accuracy of proxy benchmark using different input data.
We find that the average system and micro\comp architectural data accuracy of Proxy K-means is above 91\%  
with respect to the fully-distributed Hadoop K-means using dense input data with no zero-valued element. 
When we change the input data sparsity from 90\% to 0\%, the data accuracy of Proxy K-means is also above 
91\% with respect to the original workload. So we see that the Proxy K-means can mimic the Hadoop K-means under different input data.
The proxy benchmark accuracy will not be affected by the input data.




\subsection{Configuration Adaptability}\label{config-new}

To fulfill dynamic resource requirements in data center, the cluster might be frequently re-configured, like enlarging the memory capacity, modifying the hardware configurations, and adding more machines.
Hence, building a proxy benchmark which can suit for different configurations is of great significance. 
To evaluate the configuration adaptability of proxy benchmarks, we adopt a different cluster configuration with the cluster in Section~\ref{evaluation}, but with the same processor. 
The cluster scale is adjusted to three nodes, and the memory configuration changes to 64 GB. 


We run the original big data and AI workloads on the three-node cluster, as a comparison, we also evaluate the same proxy benchmarks on a slave node. 
The input data of all the workloads are the same as Section~\ref{evaluation}. For AI workloads, we run 3000 steps for TensorFlow AlexNet and 200 steps for TensorFlow Inception V3.

\begin{figure}[!t]
\centering
\includegraphics[scale=0.6]{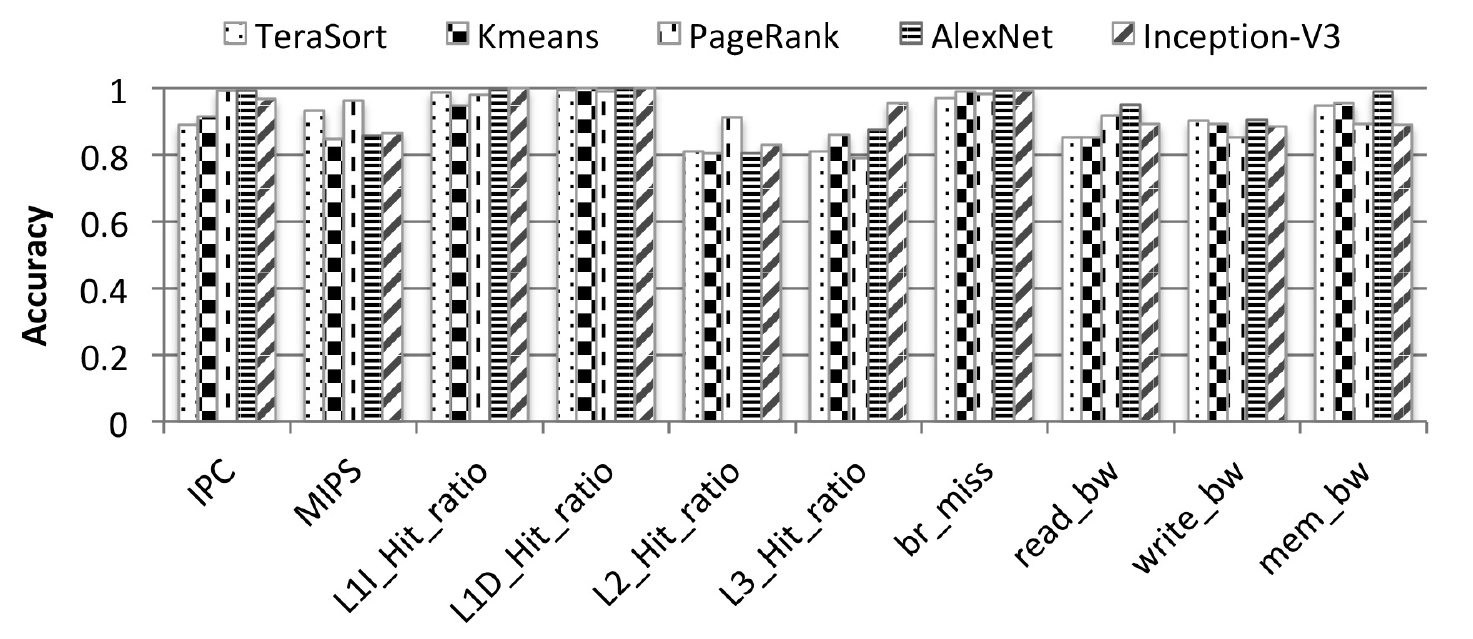}
\caption{Accuracy on a New Cluster Configuration.} 
\label{configNew:metric}
\end{figure}

\textbf{Accuracy.}
We report the system and micro-architectural data accuracy of the original workloads and the proxy benchmarks on the new cluster configuration. Likewise, we evaluate the accuracy by Equation \ref{similarity-equ1}.
Fig.~\ref{configNew:metric} presents the accuracy of the proxy benchmarks compared with the original benchmarks on the new cluster. We find that the average data accuracy of all the five proxy benchmarks are all above 90\%.
For TeraSort, K-means, PageRank, AlexNet and Inception V3, the average accuracy is 91\%, 91\%, 93\%, 94\% and 93\% respectively.

\textbf{Runtime Speedup.}
Table~\ref{time-e5645-2node} presents the execution time of the real benchmarks and the proxy benchmarks on the new cluster configuration. For big data benchmarks, Hadoop TeraSort with 100 GB text data runs 2721 seconds on the three-node cluster. Hadoop K-means with 100 GB vectors runs 7143 seconds for each iteration. Hadoop PageRank with $2^{26}$-vertex graph runs 1693 seconds for each iteration. 
For AI benchmarks, TensorFlow AlexNet with CIFAR-10 dataset runs 1333 seconds for 3000 steps. TensorFlow Inception-V3 with ILSVRC2012 dataset runs 5839 seconds for 200 steps.
The five corresponding proxy benchmarks run about twenty seconds on the physical machine. For TeraSort, K-means, PageRank, AlexNet, and Inception-V3, the speedup is 170X (2721/16.04), 509X (7143/14.03), 120X (1693/14.07), 121X (1333/11.03), and 307X (5839/19.04) respectively.

\begin{table}[htb]
\caption{Execution Time on a New Cluster Configuration.}\label{time-e5645-2node}
\renewcommand\arraystretch{1.2}
\center
\footnotesize
\begin{tabular}{|p{0.72in}|p{0.98in}|p{0.9in}|}
  \hline
  \multirow{2}{*}{Workloads} & \multicolumn{2}{c|}{Execution Time (Second)} \\
  \cline{2-3}
  & Real version & Proxy version \\
  \hline
  TeraSort & 2721 & 16.04 \\
  \hline
  K-means & 7143 & 14.03 \\
  \hline
  PageRank & 1693 & 14.07 \\
  \hline
  AlexNet & 1333 & 11.03 \\
  \hline
  Inception-V3 & 5839 & 19.04 \\
  \hline
\end{tabular}
\end{table}

\subsection{Relative Performance among Different  Architectures}

Consistent performance trend is very important for the architecture design, especially at the early stage of the system design. 
It would save a lot of time if the proxy benchmarks can be directly used on different platforms to reflect the performance of real-world workloads.
So here, we want to check whether the proxy benchmarks have the similar relative performance as the real-world workloads on different architectures.
That is to say, if the proxy benchmarks can gain the same amount of performance promotion as the real-world workloads through an improved hardware design, then the proxy benchmarks can be used to evaluate the design decision.
This subsection chooses Intel processors from different generations to reflect the different architectures and check whether the proxy benchmark can reflect the performance improvements that brought by the design of newer generation processor.  

Similar with Subsection~\ref{config-new}, we also deploy a three-node cluster, and each node is equipped with two Xeon E5-2620 v3 (Haswell) processors for performance trend comparison. The memory of each node is 64GB.
We use the proxy benchmarks to evaluate the performance trends (i.e. runtime speedup behavior) across two different architectures of Xeon E5645 (Westmere) and Xeon E5-2620 V3 (Haswell). 
All the five original workloads use the same input data on two processors, with the optimized Hadoop configurations and the same TensorFlow configurations. Meanwhile, the five proxy benchmarks are the same version and they are recompiled on the two processors.
The runtime speedup is computed using Equation \ref{speedup-time}.

\begin{equation} \label{speedup-time}
\begin{aligned}
Speedup(Time_{Westmere},Time_{Haswell})=\frac{Time_{Westmere}}{Time_{Haswell}}
\end{aligned}
\end{equation}

\begin{figure}[!t]
\centering
\includegraphics[scale=0.6]{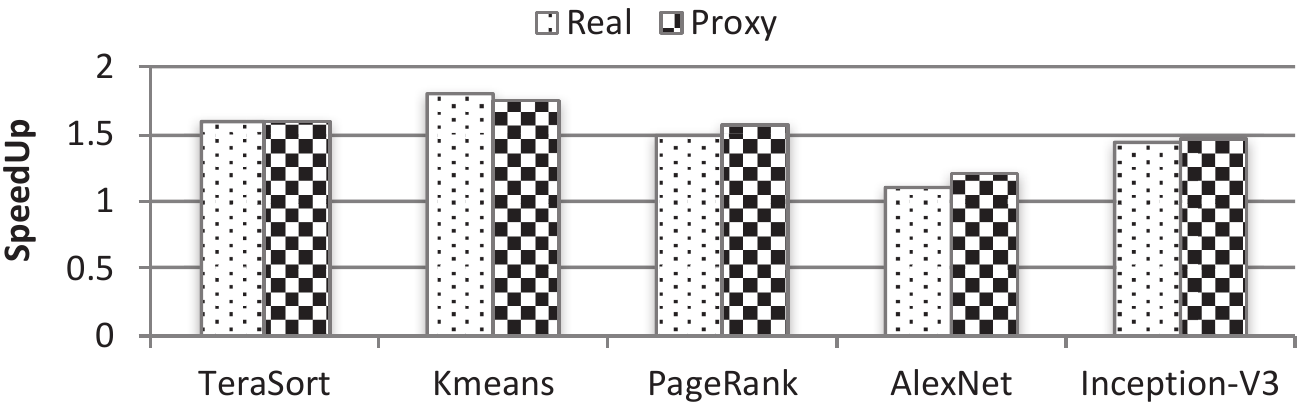}
\caption{Runtime Speedup across Westmere and Haswell Processors.} 
\label{westmere-haswell-speedup}
\end{figure}

Fig.~\ref{westmere-haswell-speedup} shows the runtime speedups of five original workloads and the proxy benchmarks across Westmere and Haswell processors. We find that for all the five proxy benchmarks, they reflect consistent speedup trends with the original big data and AI workloads. For example, the runtime speedup of Hadoop TeraSort is 1.6, running 2722 seconds on Westmere processor and 1723 seconds on Haswell processor, while the runtime speedup of Proxy TeraSort is 1.61, running 16.1 seconds on Westmere processor and 10 seconds on Haswell processor. 
Their runtime speedups range from 1.1 to 1.8 times on Haswell comparing to Westmere. AlexNet has the lowest speedup while the K-means has the highest. 

\section{Related Work}



Many research efforts about big data or AI benchmarks have been proposed in recent years.
BigBench~\cite{ghazal2013bigbench,richins2018amdahl} models a product retailer business model based on TPC-DS and provides a set of queries covering different categories of big data analytics.
BigDataBench~\cite{wang2014bigdatabench} is a benchmark suite providing dozens of big data workloads, and its latest version 4.0~\cite{gao2018bigdatabench} provides a suite of micro and component big data and AI workloads. 
CloudSuite~\cite{ferdman2011clearing} is a benchmark suite of emerging scale-out workloads and consists of eight applications. 
Also, much work about machine learning benchmarks has been presented. Fathom~\cite{adolf2016fathom} provides an AI benchmark suite consisting of eight deep learning workloads implemented with TensorFlow. BenchNN~\cite{chen2012benchnn} develops and evaluates the neural network implementations of 5 (out of 12) high performance applications from the PARSEC Benchmark Suite.
However, it is frustrating to run these benchmarks on simulators because of their complex software stacks and long running time. In addition, the simulators have limited supports for distributed deployments.

Using reduced data input is one way to reduce execution time. Previous work~\cite{kleinosowski2001adapting,kleinosowski2002minnespec} adopts reduced data set for the SPEC benchmark and maintains similar architecture behaviors with that of using the full reference data sets.  However, it does not suit for big data and AI workloads, since they are data centric computing. 


Kernel benchmarks are widely used in  high performance computing. Livermore kernels~\cite{mcmahon1986livermore} use Fortran applications to measure floating-point performance range.
 The NAS parallel benchmarks~\cite{bailey1991parallel} consist of several separate tests, including five kernels and three pseudo-applications derived from computational fluid dynamics (CFD) applications.
Linpack~\cite{dongarra2003linpack} provides a collection of Fortran subroutines. However, a single kernel is insufficient to completely reflect workload behaviors considering the complexity and diversity of emerging big data and AI workloads\cite{bailey1991parallel,lilja2005measuring}.

In terms of micro-architectural simulation, many previous studies generate synthetic benchmarks as proxies~\cite{bell2005improved,ganesan2010synthesizing}. 
Statistical simulation~\cite{skadron2003challenges,eeckhout2004control,eeckhout2000performance,nussbaum2001modeling,oskin2000hls,eeckhout2001early} generates synthetic trace or synthetic benchmarks to mimic micro-architectural performance of long-running real-world  workloads, which targets one workload on a specific architecture with the certain configurations, and thus each benchmark needs to be generated on the other architectures with different  configurations~\cite{joshi2007constructing}.
Sampled simulation selects a series of sample units for simulation instead of entire instruction stream, which are sampled randomly~\cite{conte1996reducing}, periodically~\cite{wunderlich2003smarts,yu2009tss} or based on phase behavior~\cite{sherwood2002automatically}.
Seongbeom et al.~\cite{kim2007accelerating} accelerate the full-system simulation through characterizing and predicting the performance behavior of OS services.
For emerging big data workloads, PerfProx~\cite{panda2017proxy} proposes a proxy benchmark generation framework for real-world database applications through characterizing low-level dynamic execution characteristics. The proxy benchmark of PerfProx also needs to be regenerated under different configurations.

\section{Conclusions}



In this paper, based on the data motifs that are defined as the most time-consuming units of computation performed on different initial or intermediate data, we propose a novel proxy benchmark generating methodology to generate a suite of proxy benchmarks, consisting of the DAG-like combinations of data motifs with different weights to mimic the big data and AI workloads.
Our proxy benchmarks shorten the execution time by 100s times with respect to the original benchmarks, while maintaining the average micro-architectural and system data accuracy above 90\%.
Our three case studies show that our proxy benchmarks suit for different input data, different cluster configuration and reflect consistent performance trends with original big data and AI workloads across different processors.


\section*{Acknowledgment}

This work is supported by the National Key Research and Development Plan of China (Grant No. 2016YFB1000600 and 2016YFB1000601).
The authors are very grateful to anonymous reviewers for their insightful feedback and Prof. Simone Campanoni for his  instructive suggestions. We also thank Dr. Biwei Xie for his valuable opinions.


\end{document}